\documentclass[12pt]{article}
\usepackage{geometry,amsmath,amssymb,graphicx,hyperref}
\newcommand{\mathe}{\mathrm{e}}
\newcommand{\tmem}[1]{{\em #1\/}}
\newcommand{\tmop}[1]{\ensuremath{\operatorname{#1}}}
\newcommand{\tmtextbf}[1]{{\bfseries{#1}}}
\newcommand{\tmtextit}[1]{{\itshape{#1}}}
\newcommand{\tmtexttt}[1]{{\ttfamily{#1}}}
\newcommand{\Dslash}{\ensuremath D\kern-0.65em/}

\begin{document}

\title{A factorization algorithm to compute Pfaffians}
\author{J\"urgen Rubow and Ulli Wolff\thanks{
e-mail: uwolff@physik.hu-berlin.de} \\
Institut f\"ur Physik, Humboldt Universit\"at\\ 
Newtonstr. 15 \\ 
12489 Berlin, Germany
}
\date{}
\maketitle

\begin{abstract}
We describe an explicit algorithm to factorize an even antisymmetric $N^2$
matrix into triangular and trivial factors. 
The construction resembles the Crout algorithm for LU factorization.
It allows for a 
straight forward computation of Pfaffians (including their signs)
at the cost of $N^3/3$ flops.
\end{abstract}
\begin{flushright} HU-EP-11/07 \end{flushright}
\thispagestyle{empty}
\newpage

Pfaffians play a role in statistical physics as well as in quantum field
theories (QFT) related to particle physics. They are defined for antisymmetric
even-sized quadratic matrices $A$ with elements $a_{i j} \in \mathbb{C}$ and
$i, j = 1, \ldots, N = 2 n$. As basic definition we take
\begin{equation}
  \tmop{Pf} (A) = \frac{1}{2^n n!} \sum_{\pi \in S_N} \tmop{sgn} (\pi) a_{\pi
  (1) \pi (2)} a_{\pi (3) \pi (4)} \cdots a_{\pi (N - 1) \pi (N)},
  \label{Pfdef}
\end{equation}
a sum over permutations in the symmetric group of $N$ elements reminiscent of
the definition of determinants. In the path integral formulation of QFT one
encounters Gaussian Grassmann integrals {\cite{Berezin}} for Majorana fermions
of the form
\begin{equation}
  \int D \xi \mathe^{- \frac{1}{2} \xi^{\top} A \xi} = \tmop{Pf} (A) .
  \label{Pfxi}
\end{equation}
Here the components $\xi_i, i = 1, \ldots, N$ carry indices that stand for a
compound labeling a Euclidean lattice site $x$ and a Dirac spinor component.
The identity follows from the rules of Grassmann integration with $D \xi$
being a product over all differentials ordered such that the sign works out.
In applications like {\cite{Wolff:2008xa}} the antisymmetric matrix $A
=\mathcal{C}( \Dslash + m)$ is built from charge conjugation $\mathcal{C}$ and
a (lattice) Dirac operator $\Dslash$. It may depend on other fields such as a
scalar field ($m \rightarrow m + \varphi (x)$) for the Gross Neveu model
{\cite{Korzec:2006hy}}, {\cite{Bar:2009yq}}. Majorana fermions and Pfaffians
appear almost unavoidably in formulations of supersymmetric models, see
{\cite{Campos:1999du}} for an early attempt of a lattice simulation as well as
{\cite{Catterall:2008dv}} for more recent results.

Up to a {\tmem{nontrivial}} sign a Pfaffian is the square root of a
determinant. This may be shown by doubling the Majorana fermion (\ref{Pfxi})
into a Dirac fermion
\begin{equation}
  [\tmop{Pf} (A)]^2 = \int D \xi D \xi' \mathe^{- \frac{1}{2} (\xi^{\top} A
  \xi + \xi'^{\top} A \xi')} = \int D \psi D \overline{\psi} \mathe^{-
  \overline{\psi} A \psi} = \det (A)
\end{equation}
with $\psi = (\xi + i \xi') / \sqrt{2}$, $\overline{\psi} = (\xi - i
\xi')^{\top} / \sqrt{2}$ and a corresponding definition of $D \psi D
\overline{\psi}$. If we choose the Majorana representation for the Dirac
matrices, then $A$ is manifestly real. Then there is a pseudofermion
representation for each degenerate fermion pair{\footnote{Note that this is
not suitable to simulate a single Dirac fermion, if its Majorana components
are mixed by non-real interactions.}}
\begin{equation}
  [\tmop{Pf} (A)]^2 = [\det (A^2)]^{1 / 2} = \int D \varphi \mathe^{-
  \frac{1}{2} \varphi^{\top} (A^2)^{- 1} \varphi}
\end{equation}
in terms of real bosonic variables $\varphi$ which can be a starting point of
a hybrid Monte Carlo simulation as in {\cite{Korzec:2006hy}}.

For smaller systems, in particular in two dimensions and for algorithmic
investigations, it can be of interest to compute Pfaffians and determinants
exactly in simulations {\cite{Lang:1997ib}}, {\cite{Knechtli:2003yt}} and for
other purposes {\cite{Wolff:2008xa}}, even if practicable algorithms \ cost
proportional to $N^3$. As the sign of the Pfaffian cannot be obtained from
algorithms for determinants we find it of some interest to describe in this
letter{\footnote{Based on the bachelor thesis by J.R., Humboldt university,
2009.}} an algorithm for the Pfaffian itself. Gaussian
elimination schemes suitably adapted to antisymmetric matrices are described
in {\cite{krauth2006}} and are also mentioned or contained in the deeper
layers of algorithms described in recent works requiring the computation of
Pfaffians {\cite{Catterall:2008dv}}, {\cite{PhysRevE.80.046708}},
{\cite{gonzalez2010numeric}}, {\cite{wimmer2011efficient}}. The recursive
factorization formulae worked out below are to our knowledge however not in
the literature in this easily progammable explict form.

For a nonsingular antisymmetric $A$ there exists a factorization of the form
\begin{equation}
  A = P J P^{\top} \label{PJP}
\end{equation}
where $P$ is lower triangular. The trivial antisymmetric matrix $J$ with
$\tmop{Pf} (J) = 1$ has antisymmetric $2 \times 2$ blocks \ around the
diagonal. Its nonzero elements are enumerated by
\begin{equation}
  J_{i \, i - (- 1)^i} = - (- 1)^i, \hspace{1em} i = 1, 2, \ldots N.
\end{equation}
We have learned (\ref{PJP}) in {\cite{Campos:1999du}} where it is attributed
to {\cite{Bourbaki}}. In the form of our algorithm we below give a
constructive proof for it. With the factorization given, we may change
variables in the Grassmann integral (\ref{Pfxi}) $P^T \xi = \eta$ with the
Jacobian{\footnote{Remember that the fermionic Jacobian is inverse to the
bosonic one {\cite{Berezin}}.}} $D \xi = \det (P) D \eta$ and, proving in
passing another well-known relation, we obtain
\begin{equation}
  \tmop{Pf} (A) = \det (P) \int D \eta \mathe^{- \frac{1}{2} \eta^{\top} J
  \eta} = \det (P) \tmop{Pf} (J) = \prod_{i = 1}^N p_{i i} .
\end{equation}

The factorization (\ref{PJP}) will be constructed in a way similar to the
Crout algorithm for the $L U$ factorization {\cite{Press:1058314}} of general
matrices. The matrix $P$ has more independent entries than $A$, but the
factorization is rendered unique by setting for all {\tmem{odd}} $i$
\begin{equation}
  p_{i i} = 1, \hspace{1em} p_{i + 1 i} = 0, \hspace{1em} i = 1, 3, 5, \ldots,
  N - 1 \label{condx}
\end{equation}
beside having $p_{i j} = 0$ for $i < j$.

The basic idea of the algorithm is to write out (\ref{PJP}) in components in a
special order by considering for $i = 1, 3, 5 \ldots$ the pairs
\begin{eqnarray}
  p_{j i + 1} \text{} & = & a_{i j} - {\sum_{k = 1}^{i - 2}}' (p_{i k} p_{j k +
  1} - p_{i k + 1} p_{j k}), \hspace{1em} j = i + 1, i + 2, \ldots, N, 
  \label{pa}\\
  (- p_{i + 1 i + 1}) p_{j i} \text{} & = & a_{i + 1 j} - {\sum_{k = 1}^{i -
  2}}' (p_{i + 1 k} p_{j k + 1} - p_{i + 1 k + 1} p_{j k}), \hspace{1em} 
  \label{pb}\\
  &  & \hspace{6cm} j = i + 2, i + 3, \ldots, N. \nonumber
\end{eqnarray}
Here the primed sums over $k$ run over {\tmem{odd integers only}} and
(\ref{condx}) has been exploited. If we assume for a moment that we can always
divide by $p_{i + 1 i + 1}$ in (\ref{pb}) then we may column-wise solve for
the nontrivial $p_{j 2}, p_{j 1}, p_{j 4}, p_{j 3}, \ldots p_{N N}$. It is
decisive to note that in this order we only encounter columns of $P$ on the
right hand sides of (\ref{pa}), (\ref{pb}) that have been computed before.

Even for simple regular matrices the assumption $p_{i + 1 i + 1} \neq 0$ may
not always be fulfilled in the steps with (\ref{pb}). Both for this reason and
to improve the numerical precision a pivoting scheme is mandatory. To that
end we introduce a permutation $\pi$ as in (\ref{Pfdef}) and replace $A, P$ by
re-arranged copies $A', P'$ with matrix elements
\begin{equation}
  a'_{i j} = a_{\pi (i) \pi (j)}, \hspace{1em} p'_{i j} = p_{\pi (i) j}
\end{equation}
and note that (\ref{PJP}) is equivalent to $A' = P' J P'^{\top}$and thus
`covariant' under such a relabelling. We consider the above process now for
$A', P'$ with $\pi$ initially set to the identity. Each time after completing
(\ref{pa}) for some $i$ we now determine the value $j$ along the column where
$|p'_{j i + 1} |$ is maximal and denote it by $j_{\max}$. Before proceeding
with (\ref{pb}) we swap the entries $\pi (i + 1) \leftrightarrow \pi (
\text{$j_{\max}$})$. It is important to note that this modification does not
invalidate{\footnote{This is because we have generated complete columns whose
entries are permuted in the same way on both sides of the equations. The
column indices of earlier columns on the other hand are untouched as $j_{\max}
\geqslant i + 1$.}} any of the earlier uses of (\ref{pa}) and (\ref{pb}). In
this way we never divide by zero except when 
the whole column constructed via
(\ref{pa}) vanishes, in which case one can show that $A$ is singular.
In all
other cases we arrive at the factorization of the matrix $A' = \Pi A
\Pi^{\top}$ where the matrix $\Pi$ implements the index permutation with
$\pi$. Then we obtain $\tmop{Pf} (A)$ from
\begin{equation}
  \tmop{Pf} (A) = \det (\Pi^{- 1}) \tmop{Pf} (A') = \tmop{sgn} (\pi) \prod_{i
  = 1}^n p'_{2 i 2 i} .
\end{equation}
Our $P J P$ factorization for computing the Pfaffian requires
approximately $N^3/3$ flops (counting both  multiplications and additions). The signum
is known by counting the transpositions that went into building $\pi$.

\begin{figure}[htb]
\begin{center}
  \resizebox{0.7\textwidth}{!}{\includegraphics{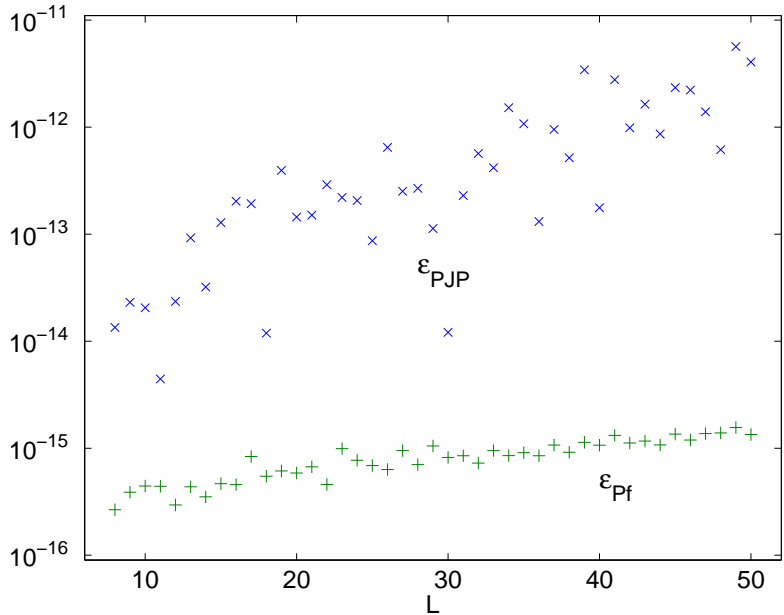}}
  \caption{Precision of $\tmop{Pf}$ ($\times$) and the factorization ($+$)
  defined in (\ref{devs}) as functions of the lattice size $L$ associated
  with matrices of size $N=2L^2$.\label{fig}}
\end{center}
\end{figure}

We end by reporting on a brief test. We consider the Wilson fermion matrix on
an $L \times L$ square lattice
\begin{equation}
  A_W = c_0 \mathcal{C} \left[ 2 - \frac{1}{2} \sum_{\mu = 0, 1} \{(1 -
  \gamma_{\mu}) \delta_{x, x - \hat{\mu}} + (1 + \gamma_{\mu}) \delta_{x, x +
  \hat{\mu}} \} \right] \label{AW}
\end{equation}
which is an antisymmetric matrix with $N = 2 L^2$. The Dirac matrices are
given in terms of Pauli matrices, $\mathcal{C}= i \tau_2$, $\mathcal{C}
\gamma_0 = \tau_1$, $\mathcal{C} \gamma_1 = \tau_3$, $\hat{\mu}$ is a unit
vector in the positive $\mu$ direction, and the $\delta$ symbols here
incorporate {\tmem{antiperiodic}} boundary conditions in both directions thus
making $A_W$ nonsingular without a mass term. The Pfaffian for this matrix can
be computed exactly by Fourier transformation {\cite{Wolff:2007ip}}. To avoid
a range overflow we have used this information to determine $c_0$ in (\ref{AW}) such that
$\tmop{Pf} (A_W) = 1$ holds up to roundoff errors. We have coded the procedure
described above and have run it with $L = 8, 9, \ldots, 50$ in standard 64 bit
precision. In Figure \ref{fig} we plot the deviations
\begin{equation}
  \varepsilon_{\tmop{Pf}} = | \tmop{Pf} (A_W) - 1|, \hspace{1em}
  \varepsilon_{\tmop{PJP}} =\|A'_W - P' J {P'}^{\top} \| \label{devs}
\end{equation}
as functions of $L$ where the matrix norm in $\varepsilon_{\tmop{PJP}}$ is
given by the largest singular value. We note that apart form a general trend
they sometimes are
exceptionally small leading to some non-uniformity with $L$.

\end{document}